\documentclass[conference]{IEEEtran}
\IEEEoverridecommandlockouts
\usepackage{cite}
\usepackage{amsmath,amssymb,amsfonts}
\usepackage{algorithmic}
\usepackage{graphicx}
\usepackage{textcomp}
\usepackage{xcolor}
\usepackage{hyperref}
\usepackage{booktabs}
\usepackage{enumerate}
\usepackage[compact]{titlesec}
\titlespacing{\section}{0pt}{1ex}{0.5ex}
\titlespacing{\subsection}{0pt}{0.5ex}{0ex}
\titlespacing{\subsubsection}{0pt}{0.5ex}{0ex}
\def\BibTeX{{\rm B\kern-.05em{\sc i\kern-.025em b}\kern-.08em
    T\kern-.1667em\lower.7ex\hbox{E}\kern-.125emX}}
\begin{document}

\title{%PCE and Global Sensitivity Indicators of a 33-bus MG
A Data-Driven Polynomial Chaos Expansion-Based Method for Microgrid Ramping Support Capability Assessment and Enhancement  
\thanks{This work was supported by Natural Sciences and Engineering Research Council (NSERC) Discovery Grant, NSERC RGPIN-2022-03236.}
}

\author{\IEEEauthorblockN{Mohan Du and Xiaozhe Wang}
\IEEEauthorblockA{Department of Electrical and Computer Engineering, 
McGill University, 
Montreal, QC H3A 2K6, Canada \\
mohan.du@mail.mcgill.ca, xiaozhe.wang2@mcgill.ca}

}

\maketitle

\begin{abstract}
%The high penetration of renewable energy sources (RES) and electric vehicles (EVs) to electric grids significantly changes the traditional vertically integrated power systems and may enhance the power unbalance in utility grids. 
Microgrids (MGs) are regarded as effective solutions to provide ramping support to the main grid during heavy-load periods. Nevertheless, the uncertain renewable energy sources (RES) and electric vehicles (EVs) integrated into an MG may affect the ramping support capability (RSC) of an MG. To address the challenge, this paper develops a data-driven sparse polynomial chaos expansion (DDSPCE)-based method to accurately and efficiently evaluate the hour-by-hour RSC of an MG. The DDSPCE model is further exploited to identify the most influential random inputs, based on which a scheduling method of BESS is developed to enhance the RSC of an MG. Simulation results in the modified IEEE 33-bus MG shows that the proposed method takes less than 3 minutes for evaluating and enhancing the hourly RSC. 
%increase the ramping support capability (RSC) of a microgrid (MG) so that the MG can provide ramping support to the utility grid during the heavy load period. 
%The developed method employed Sobol' indices to identify the most influential random inputs in the MG and scheduled BESS to smooth them out. Once the influential random inputs are smoothed, the variance of the RSC is reduced, and the RSC is increased. 
%Since the employed Sobol' indices are based on the data-driven sparse polynomial chaos expansion (DDSPCE), the developed method can identify the most significant influencers quickly without any biased assumptions of random distributions. The time efficiency of the developed method enables it to be deployed online for real-time RSC enhancement. 
%The feasibility and validity of the developed method are tested on a modified IEEE 33-bus test grid. 
\end{abstract}


\section{Introduction}
Integrating renewable energy sources (RES), such as photovoltaics (PVs) and wind turbines (WTs) to power systems is promising to decelerate global warming and achieve net-zero emission by 2050 \cite{Majzoobi2017}. However, the advantages of RES are accompanied by variability, which poses profound challenges to the operation of traditional vertically integrated power systems \cite{Chiang2015}. According to the report \cite{Denholm2013} by California Independent System Operator in 2013, the discrepancy between the RES's generation and customer's demand profiles will intensify the supply-load unbalance and result in a steep net-load ramp. The transmission system operators urgently require the service of fast-response power generation during heavy-load periods  \cite{Majzoobi2017}.

This service of fast-response power generation can be provided by microgrids (MGs), which are specific distribution grids integrated with dispatchable, e.g., diesel generators (DGs) and battery energy storage systems (BESS), and non-dispatchable, e.g., distributed generators powered by RES. Indeed, the MG is considered a viable and effective solution to provide ancillary services to main grids for improving stability, resiliency, and reliability \cite{Kumar2019}. Despite the categorization of ancillary services varies from country to country, essential ancillary services widely adopted by large power systems include voltage-VAr control, frequency-Watt control, load shedding, black start, and ramping support \cite{Kumar2019,Majzoobi2017}.

%\color{black}
Particularly, this paper considers the capability of MGs to provide ramping support during heavy-load periods. The ramping support capability (RSC) is defined as the available active power that the corresponding MG can transfer to the connected grid during a certain period \cite{Majzoobi2017,Dong2010,NorthAmericanElectricReliabilityCouncil1996,Chiang2015}. Nonetheless, the RSC of an MG may also be affected by the variability of RES and electric vehicles (EVs) inside an MG. In \cite{Majzoobi2016}, the RSC is determined by a min-max problem, aiming to calculate a worst-case ramping power at every time interval during a day. Based on the calculated RSC,  an optimization model was developed to coordinate the MG loads to settle the intense ramping issue. Similarly, various optimization models for energy scheduling and management were developed in \cite{Kumar2019,Majzoobi2016,Majzoobi2017}, in which the time-varying property of RES and/or loads are described by pre-assumed profiles. The authors of \cite{Yan2017,Alharbi2015} employed probabilistic forecasting models to predict the profiles of RES and load and quantified the size of the spinning reserve from the predicted errors by assuming a certain risk level. The authors of \cite{Wang2011a} proposed an estimation model of spinning reserve in MGs, in which the uncertainty of WTs, PVs, and loads are aggregated to reduce the computational burden. However, the power flow constraints and other security constraints such as voltage and thermal limits were not considered in \cite{Yan2017,Alharbi2015,Wang2011a}. To incorporate the security constraints and quantify the impacts of uncertainties of RES, a sparse polynomial chaos expansion (SPCE)-based method was developed in \cite{Sheng2018} to estimate the available delivery capability of a distribution system accurately and efficiently. Nevertheless, the applied method requires accurate marginal distributions of random inputs that may not always be available in practice. Besides, control measures to increase the available delivery capability were not discussed. 
\color{black}

In this paper, we will leverage the data-driven SPCE (DDSPCE) method proposed in \cite{Wang2021} to accurately and efficiently estimate the RSC of an MG considering the uncertainties of RES and EVs as well as the security constraints of an MG. Particularly, %\color{red} unlike what \cite{Li2019a,Sheng2018} did, \color{black} 
the method requires no knowledge of marginal distributions of WTs, PVs, loads, etc. It should be noted that the DDSPCE was not exploited to design control measures in \cite{Wang2021}. In contrast, in this paper, the established DDSPCE model will be further used to calculate the Sobol' indices that can identify the dominant random inputs, based on which control measures utilizing BESS are developed to increase the quality of the RSC of an MG.  Simulation results in a modified IEEE 33-bus MG integrating PVs, WTs, and EVs show that the developed method can increase the quality of the RSC of the MG significantly. 

The rest of the paper will be organized as follows. Section \ref{sec:RSC} introduces the concept and formulations of probabilistic RSC. Section \ref{sec:DDSPCE} introduces the formulations of DDSPCE and DDSPCE-based Sobol' indices. 
Section \ref{sec:steps} introduces the developed method to increase the RSC. 
Section \ref{sec:results} validates the effectiveness of the developed RSC-enhancement method in the test grid.
Section \ref{sec:conclusion} presents the conclusion.

\section{Probabilistic RSC}\label{sec:RSC}
In this paper, the continuous power flow for an $N$-bus power system is used to calculate the RSC of an MG
\begin{equation}\label{equ:cpf}
f(\boldsymbol{\varphi}, \lambda)=f(\boldsymbol{\varphi})-\lambda \boldsymbol{b}=0,
\end{equation}
where $f(\boldsymbol{\varphi})$ is the solution to the power-flow equation $f(\boldsymbol{\varphi})=0$ \cite{Sheng2018}, the state vector $\boldsymbol{\varphi}=[\boldsymbol{\theta}^{\top}, \boldsymbol{V}^{\top}]^{\top}$, $\boldsymbol{\theta}\in \mathbb{R}^{N}$ and $\boldsymbol{V}\in \mathbb{R}^{N}$ are voltage angle and magnitude vectors, respectively, and $\lambda \in \mathbb{R}$ %is a parameter indicating 
refers to the RSC of an MG. 
The %load-generation variation 
vector $\boldsymbol{b}=[\boldsymbol{b}_1^{\top}, \boldsymbol{b}_2^{\top}, \ldots, \boldsymbol{b}_{N}^{\top}]^{\top} \in \mathbb{R}^{{2N}\times 1}$ describes the direction of power transfer variation:  %i.e., from an MG to the connected utility bus. 
%indicates the direction of the load changing:
\begin{equation}\label{equ:cpf}
    \begin{aligned}
%        \boldsymbol{b}&=[\boldsymbol{b}_1^{\top}, \boldsymbol{b}_2^{\top}, \ldots, \boldsymbol{b}_{N}^{\top}]^{\top}, \\
        \boldsymbol{b}_1&=\left[\begin{array}{c}
        -\Delta P_{L, 1} \\
        -\Delta Q_{L, 1}
        \end{array}\right], \\
         \boldsymbol{b}_i&=\left[\begin{array}{c}
        \Delta P_{G, i}-\Delta P_{L, i} \\
        -\Delta Q_{L, i}
        \end{array}\right], \quad i=2, 3, \ldots, N,
    \end{aligned}
\end{equation}
where $\boldsymbol{b}_1$ is a unit vector, i.e., $\|\boldsymbol{b}_1\|_2=1$, %the unit vector $\|\boldsymbol{b}_1\|_2=1$ indicates the direction of the ramping support, 
bus 1 is the connected utility bus; %where the power transfer from an MG wants to reach; 
$\Delta P_{G, i}$ is the assigned increase of active generation power from the MG; $\Delta P_{L, i}$, and $\Delta Q_{L, i}$ are the increase of active load, and reactive load inside the MG, respectively. Particularly, bus 1 is modeled as a PQ load bus such that the %dispatchable distributed energy sources 
the real power of an MG will transfer power to the main grid.  

Considering the uncertainties of RES and EVs in the MG, the continuous power flow equation \eqref{equ:cpf} can be modified to the probabilistic continuous power flow equation \cite{Sheng2018}, which reads 
\begin{equation}\label{equ:pcpf}
f(\boldsymbol{\varphi}, \lambda, \boldsymbol{x})=f(\boldsymbol{\varphi},\boldsymbol{x})-\lambda \boldsymbol{b}=0,
\end{equation}
where the random vector $\boldsymbol{x}$ describes the random inputs, i.e., solar radiation, wind speed, and the charging power of EVs, that affect the power generations and loads in the MG. 
The formulation of the probabilistic RSC reads:
\begin{equation}\label{equ:pADC}
    \begin{array}{ll}
        \max \lambda & \\
        \text {s.t.} & \boldsymbol{f}(\boldsymbol{\varphi}, \boldsymbol{x})-\lambda \boldsymbol{b}=0 \\
        & V_{\min } \leq V_i(\boldsymbol{\varphi}, \lambda, \boldsymbol{x}) \leq V_{\max }\\
        & I_{i j}(\boldsymbol{\varphi}, \lambda, \boldsymbol{x}) \leq I_{i j, \max } \\
        & P_{\min , i} \leq P_{G i}(\boldsymbol{\varphi}, \lambda, \boldsymbol{x}) \leq P_{\max , i},\\
        & Q_{\min , i} \leq Q_{G i}(\boldsymbol{\varphi}, \lambda, \boldsymbol{x}) \leq Q_{\max , i},\\
        & i, j \in\{1, 2, \ldots, N\},
\end{array}
\end{equation}
where $V_{\max }$ and $V_{\min }$ are upper and lower limits of bus voltages, respectively, $I_{i j, \max }$ is the thermal limit of line $ij$, $P_{\max,i }$ and $P_{\min,i }$ are the maximum and minimum output powers of the generator on bus $i$, respective, which applies to $Q_{\max,i }$ and $Q_{\min,i }$ similarly.
The maximum $\lambda$ without violating any constraint in \eqref{equ:pADC} is the RSC of an MG.  %margin under specified load-generation vector $\boldsymbol{b}$. 

It should be noted that $\lambda$ is a random variable because of the random input $\boldsymbol{x}$ in \eqref{equ:pADC}. Once the MG configuration and $\boldsymbol{b}$ are determined,  $\lambda(\boldsymbol{x})$ can be described as a function of $\boldsymbol{x}$ according to \eqref{equ:pADC}. 
%\color{black}
The traditional method to estimate the distribution of $\lambda(\boldsymbol{x})$ is to perform Monte Carlo simulations (MCS) on \eqref{equ:pADC}. 
However, whatever efficient sampling method, e.g., Latin hypercube \cite{Yu2009} or importance sampling \cite{Huang2011}, is used, the MCS is inevitably computationally expensive \cite{Sheng2018}. 
To overcome the problem of time consumption, the DDSPCE-based method \cite{Wang2021} is developed. 
\color{black}
% descriptions needed that why we need to use DDSPCE to estimate $\lambda$. connecting sentences between different sections are needed to keep the flow.

\section{DDSPCE and DDSPCE-Based Sobol' Indices}\label{sec:DDSPCE}

The DDSPCE method aims to use a sparse finite degree model $\hat{\lambda}=g(\boldsymbol{x})$ to approximate a stochastic model, e.g., (\ref{equ:pADC}),  %$\lambda=f(\boldsymbol{x})$ 
with a target stochastic response $\lambda\in \mathbb{R}$ and a random input vector $\boldsymbol{x}=[x_1, x_2, \ldots, x_D]^{\top}$ \cite{Liu2022,Marelli2022,Wang2021}.
The DDSPCE method can achieve both time efficiency and accuracy by building the stochastic model $\hat{\lambda}=g(\boldsymbol{x})$ using only %$\hat{\lambda}$ with 
a few sample pairs of $(\boldsymbol{x}, \lambda)$.  %and corresponding inputs $\boldsymbol{x}$. 
Mathematcially, the DDSPCE model $\hat{\lambda}=g(\boldsymbol{x})$ can be described as: %is defined as: 
\begin{equation}\label{equ:pce}
\begin{aligned} 
    \hat{\lambda} & = g(\boldsymbol{x})=\sum_{\alpha \in \mathcal{A}} c_{\alpha} \Psi_{\alpha}(\boldsymbol{x}),
\end{aligned}
\end{equation}
where the 1-norm truncated set $\mathcal{A}=\{\alpha\in\mathbb{N}^D : \|\alpha\|_1\leq q\}$, $q$ is the degree of truncation, $\alpha\in \mathcal{A}$ is an $N$-dimensional index, and $c_{\alpha}\in\mathbb{R}$ and $\Psi_{\alpha}(\boldsymbol{x})$ are the coefficient and polynomial basis corresponding to $\alpha$, respectively \cite{Marelli2022}. 
$\Psi_{\alpha}(\boldsymbol{x})$ is calculated as \cite{Wang2021}:
\begin{equation}\label{eq:mutibasis}
\Psi_{\alpha}\left(x_1, \ldots, x_{D}\right)=\prod_{i=1}^{D} \left( \sum_{k=0}^{\alpha_i} p_{\alpha_i,k}x_i^k \right). 
\end{equation}
The univariate polynomial basis $p_{\alpha_{i},k}$ is the solution to
\begin{equation}
    \begin{aligned}
    & \left[\begin{array}{cccc}
        \mu_{0, i} & \mu_{1, i} & \ldots & \mu_{\alpha_{i}, i} \\
        \mu_{1, i} & \mu_{2, i} & \ldots & \mu_{\alpha_{i}+1, i} \\
        \vdots & \vdots & \vdots & \vdots \\
        \mu_{\alpha_{i}-1, i} & \mu_{\alpha_{i}, i} & \ldots & \mu_{2 \alpha_{i}-1, i} \\
        0 & 0 & \ldots & 1
        \end{array}\right]\left[\begin{array}{c}
        p_{\alpha_i,0} \\
        p_{\alpha_i, 1}\\
        \vdots \\
        p_{\alpha_i, \alpha_{i}-1} \\
        p_{\alpha_{i}, \alpha_i}
        \end{array}\right]=\left[\begin{array}{c}
        0 \\
        0 \\
        \vdots \\
        0 \\
        1
        \end{array}\right], 
    \end{aligned}
\end{equation}
where  $\mu_{k, i}=\int_{x_i \in \Omega_i} x_i^k d \Gamma\left(x_i\right)$ is the raw moment of $x_i$, and $\Omega_i$ is the space set of $x_i$.  
Particularly, $\mu_{k, i}$  can be obtained from historical/predicted data or probabilistic models of $x_i$. 

Once $p_{\alpha_{i},k}$ is solved, the multivariate polynomial basis $\Psi_{\alpha}(\boldsymbol{x})$ can be built from (\ref{eq:mutibasis}), the remaining task is to calculate the coefficients $c_{\alpha}$, %The coefficients $c_{\alpha}$ 
which can be obtained by some advanced regression methods, e.g., the ordinary least-square method \cite{Wang2021}:
\begin{equation}\label{equ:pce-c}
    \boldsymbol{C}=\left(\boldsymbol{\Psi}(\boldsymbol{X})^{\top} \boldsymbol{\Psi}(\boldsymbol{X})\right)^{-1} \boldsymbol{\Psi}(\boldsymbol{X})^{\top} \boldsymbol{\lambda},
\end{equation}
where $\boldsymbol{X}=[\boldsymbol{x}_{1}, \boldsymbol{x}_{2}, \ldots, \boldsymbol{x}_{N}]^{\top}\in \mathbb{R}^{N}\times\mathbb{R}^{D}$ and $\boldsymbol{\lambda}=[\lambda_{1}, \lambda_{2}, \ldots, \lambda_{N}]^{\top}\in \mathbb{R}^{N}$ are random inputs and stochastic responses used for the regression, $\boldsymbol{C}=[\ldots, c_{\alpha},\ldots]^{\top}\in \mathbb{R}^{N_{\alpha}}$ and $\boldsymbol{\Psi}(\boldsymbol{X})=[\ldots, \Psi_{\alpha}(\boldsymbol{X}),\ldots]\in \mathbb{R}^{N}\times\mathbb{R}^{N_{\alpha}}$ are matrices of $c_{\alpha}$ and $\Psi_{\alpha}(\boldsymbol{X})$, respectively, $\Psi_{\alpha}(\boldsymbol{X})=[\Psi_{\alpha}(\boldsymbol{x}_1), \Psi_{\alpha}(\boldsymbol{x}_2), \ldots, \Psi_{\alpha}(\boldsymbol{x}_N)]^{\top}\in \mathbb{R}^{N}$, 
and $N_{\alpha}$ is the number of elements in $\mathcal{A}$. 

% For the sake of simplicity, all $\mathcal{A}^{N,D}$ is simplified as $\mathcal{A}$ if there is no specific explanation. 

After the calculation of $c_\alpha$, the DDSPCE model (\ref{equ:pce}) is built. Next, the Sobol' index of each random variable can be calculated. 
The Sobol' index $S_i$ quantifies the effect of $x_i$ on the variance of $\hat{\lambda}$ \cite{Marelli2022a}. A larger $S_i$ indicates that $x_i$ plays a more important role in affecting $\mathrm{Var}[\hat{\lambda}]$. For this reason, Sobol' indices can be used to identify the dominant influencers among all uncertainty sources. %that contribute to $\mathrm{Var}[\hat{\lambda}]$ significantly.\color{black}
%\color{red} explain what Sobol' indices mean. \color{black}
The Sobol' decomposition is defined as \cite{Marelli2022a}
\begin{equation}
\begin{aligned}
\hat{\lambda}=g(\boldsymbol{x}) 
% & = g_0+\sum_{i_1=1}^n g_{i_1}\left(X_{i_1}\right)+\sum_{1 \leq i_{1}<i_{2} \leq n} g_{i_1, i_2}\left(x_{i_1}, x_{i_2}\right)+\ldots\\
% & + \sum_{1 \leq i_1<\ldots<i_k \leq n} g_{i_1, 2, \ldots, i_k}\left(x_{i_1}, \ldots, x_{i_k}\right)+\ldots + g_{1, 2, \ldots, n}\left(x_1, 2, \ldots, x_n\right) \\
% 
=g_0+\sum_{\substack{\boldsymbol{u} \subseteq\{1, 2, \ldots, D\} \\ \boldsymbol{u}\neq \varnothing}} g_{\boldsymbol{u}}\left(\boldsymbol{x}_{\boldsymbol{u}}\right),
\end{aligned}
\end{equation}
where for any non-empty set $\boldsymbol{u}\subseteq\{1, 2, \ldots, D\}$, $\boldsymbol{x}_{\boldsymbol{u}}= \{x_i\in \boldsymbol{x}| i\in \boldsymbol{u} \}$ and $g_{\boldsymbol{u}}(\boldsymbol{x}_{\boldsymbol{u}})$ is a function of $\boldsymbol{x}_{\boldsymbol{u}}$. 

We can easily perform Sobol' decomposition on \eqref{equ:pce} by defining \cite{Marelli2022a}
\begin{subequations}\label{equ:sobol_modify}
\begin{align}
    \label{equ:guxu} &g_{\boldsymbol{u}}(\boldsymbol{x}_{\boldsymbol{u}})=\sum_{\alpha\in\mathcal{A}_{\boldsymbol{u}}}c_{\alpha} \Psi_{\alpha}(\boldsymbol{x}),\\
    & \mathcal{A}_{\boldsymbol{u}}=\{\alpha\in \mathcal{A} | \alpha_u\neq 0 \Leftrightarrow u\in \boldsymbol{u}, u=1, 2, \ldots, D\}. 
\end{align}
\end{subequations}
Equation \eqref{equ:guxu} holds because the basis $\Psi_{\alpha}(\boldsymbol{x})$ is irrelevant to random variables $x_i\in \{x_i\in \boldsymbol{x}| \alpha_i=0\}$.
According to \eqref{equ:sobol_modify}, \eqref{equ:pce} can be re-expressed as 
\begin{equation}
    \hat{\lambda}  = g(\boldsymbol{x})= c_0 +
    \sum_{\substack{\boldsymbol{u} \subseteq\{1, 2, \ldots, D\}\\ \boldsymbol{u}\neq \varnothing}}
    \sum_{\alpha \in \mathcal{A}_{\boldsymbol{u}}} c_{\alpha} \Psi_{\alpha}(\boldsymbol{x}), 
\end{equation}
where $c_0$ is the expectation of $\hat{\lambda}$. 

Considering the orthonormality of polynomial chaos bases \cite{Marelli2022}, the variance of the DDSPCE model reads: %\color{red} use  $\mathrm{Var}$ \color{black} fixed \color{black} 
\begin{equation}
\begin{aligned}
\mathrm{Var}\left[\hat{\lambda}\right] &=\sum_{\substack{\alpha \in \mathcal{A} \\
\alpha \neq 0}} \widehat{c}_{\alpha}^2, \\
\mathrm{Var}\left[g_{\boldsymbol{u}}\left(x_{\boldsymbol{u}}\right)\right] &=\sum_{\substack{\alpha \in \mathcal{A}_{\boldsymbol{u}} \\
\alpha \neq 0}} \widehat{c}_{\alpha}^2. 
\end{aligned}
\end{equation}
The Sobol' index of $\boldsymbol{x}_{\boldsymbol{u}}$ is expressed as 
\begin{equation}\label{equ:sobol}
    S_{\boldsymbol{u}}=\frac{\sum_{\substack{\alpha \in \mathcal{A}_{\boldsymbol{u}} \\
\alpha \neq 0}} \widehat{c}_{\alpha}^2}
{\sum_{\substack{\alpha \in \mathcal{A} \\
\alpha \neq 0}} \widehat{c}_{\alpha}^2}.
\end{equation}

% Specifically, if $\boldsymbol{u}=\{1\}$, $S_{1}$ describes the contribution of $x_1$ on $\mathrm{Var}[\hat{\lambda}]$. %$\hat{\lambda}=g(\boldsymbol{x})$. 
% The larger $S_{\boldsymbol{u}}$ is, the higher variance $\boldsymbol{x}_{\boldsymbol{u}}$ contributes to $\mathrm{Var}[\hat{\lambda}]$.

\section{The DDSPCE-based Algorithm for RSC Assessment and Enhancement}\label{sec:steps}%Calculation of RSC and Identification of Influencers}
\label{sec:calculatingRSC}
Due to the uncertainty of random inputs, the RSC of an MG may have a large variance which indicates a worse quality. 
This section utilizes the DDSPCE-based Sobol' indices to identify the influential random inputs and smooth out their outputs with BESS. 
Once the outputs of the most influential random inputs are smoothed, the variance of the RSC is reduced, and the RSC quality is increased. 
The detailed steps of the proposed RSC-enhancement algorithm are provided below. 
\begin{enumerate}[Step 1:]
\item \label{step:generate} %Randomly 
Acquire $N_{0}$ samples of $D$ random inputs $\boldsymbol{x}$ (e.g., wind speed, solar radiation, and EV power) from historical/predicted data or probabilistic models. The samples are denoted as a $N_0\times D$ matrix $\boldsymbol{X}_{0}=[\boldsymbol{x}_{1}, \boldsymbol{x}_{2}, \ldots, \boldsymbol{x}_{N_0}]^{\top}$, where each row and column correspond to a sample and random variable, respectively. 
%for each month $m$ and hour $h$, where $m=1, \ldots, 12$ and $h=1, \ldots, 24$ are month and hour indices, respectively. The acquired random input for pair $(m,h)$ is denoted by $\mathbf{X}_{0}^{m,h}=[\boldsymbol{x}^{m,h}_{0,1}, \boldsymbol{x}^{m,h}_{0,2}, \ldots, \boldsymbol{x}^{m,h}_{0,N}]^{\top}$ is a $N\times D$ matrix, where each row and column correspond to a scenario and random variable, respectively.
\item Calculate the corresponding RSC $\boldsymbol{\lambda}_{0}=[\lambda_{1}, \lambda_{2}, \ldots, \lambda_{N_0}]^{\top}$ by solving \eqref{equ:pADC}, where each $\lambda_{i}$ corresponds to the input $\boldsymbol{x}_{i}$. 
%$\lambda_0$ for these samples  $\boldsymbol{x_0}$ %\boldsymbol{y}_{0}=[y^{m,h}_{0,1}, y^{m,h}_{0,2}, \ldots, y^{m,h}_{0,N}]^{\top}$ for each $(m,h)$ 
%by solving \eqref{equ:pADC}. %where each $y^{m,h}_{0,i}$ corresponds to the input $\boldsymbol{x}^{m,h}_{0,i}$. 
\item Build the DDSPCE model $\hat{\lambda}=g(\boldsymbol{x})$ based on the sample pairs $(\boldsymbol{X}_{0}, \boldsymbol{\lambda}_{0})$ according to \eqref{equ:pce}-\eqref{equ:pce-c}. %You have to at least explain how to build the model in this paper. Simply referring [20] is not good practice
\item \label{step:RSC} Substitute $N_s\gg N_0$, a large number of samples of $\boldsymbol{x}$, i.e.,  $\boldsymbol{X_s}$,  to the established DDSPCE model to calculate the corresponding $\boldsymbol{\hat{\lambda}_s}$, i.e., $\boldsymbol{\hat{\lambda}}_{s}=g(\boldsymbol{X}_{s})$. From the probability distribution function (PDF) of $\boldsymbol{\hat{\lambda}_{s}}$ and a given a confidence level $\gamma\%$ (e.g., $95\%$), the RSC of an MG with a confidence level $\gamma\%$ can be estimated by $P(RSC>\hat{\lambda})=\gamma\%$. %where the confidence level $0<\gamma\%<1$ is generally set to be $95\%$. Since the larger $N_s$ is, the more accurate $RSC^{m,n}$ can be obtained, $N_s$ is generally set to be a large number. 
\item \label{step:dominant} Calculate the Sobol' index $S_{i}$ of each random variable $x_i$ by \eqref{equ:sobol} and identify the top $N_{b}$ random variables such that $\sum_{i=1}^{N_{b}}S_{i}\geq 80\%$. %\color{black} later \color{black} %where $N_{b}$ is the number of BESS in the MG. 

\item Implement $N_{b}$ BESS at the buses where the dominant random variables locate to smooth out the output powers,   %whose RES or load contribute most to $\mathrm{Var}[\hat{\lambda}^{m,h}]$ and use BESS 
i.e., reduce the variances of dominant random variables' output powers 
% corresponding to the dominant random variables identified in \textbf{Step 5} 
to zero.  
%We developed two types of plans to decide the location of BESS. 
%For the first one, four BESS are installed on the buses where the four selected variables are corresponding to. 
%For the second one, four BESS are installed on the buses corresponding to the largest Sobol' index on each branch. 
Specifically, for each BESS $i, i=1, \ldots, N_b$ at bus $b_i$, %corresponding to the dominant random variable $x_i$, 
the output power $ P_{B,i}$ of BESS $i$ is determined by 
%\color{red} $i$ or $b_i$, be consistent. \color{black}
%\color{black} Hi professor, in this equation, $i$ is the index of the BESS, and $b_i$ is the index of the bus where BESS $i$ locates. For example, our BESS 1 ($i=1$) locates on bus $7$ ($b_i=7$) in Fig. \ref{fig:test_grd}.  \color{black}
\begin{equation}\label{equ:Pbesssmooth}
    \begin{aligned}
    \min_{P_{B,i}} \, &  |P_{B,i}-(\mathbb{E}[P_{G,b_i}]-P_{G,b_i}-\mathbb{E}[P_{L,b_i}]+P_{L,b_i})|, \\
    \text{s.t. } & P_{\min, B,i}\leq P_{B,i} \leq P_{\max, B,i}, \\
    & 0\leq SOC_{B,i} \leq SOC_{\max, B,i},\\
    & i=1, 2, \ldots, N_b
    \end{aligned}
\end{equation}
where %$b_i$ is the index of the bus where BESS $i$ and corresponding dominant random input locate, 
$P_{\min, B,i}$ and $P_{\max, B,i}$ are the %universal 
minimum and maximum output power of BESS $i$, %installed in the MG, 
$SOC_{B,i}$ is the state of charge of  BESS $i$, and $SOC_{\max, B, i}$ is the capacity of BESS $i$. 
\end{enumerate}

Note that once the DDSPCE model is built, the evaluation of $\boldsymbol{\hat{\lambda}_s}$ in \textbf{Step 4} takes negligible time as the DDSPCE model is a simple algebraic model that is very fast to evaluate compared to the original model (\ref{equ:pADC}). That is how the DDSPCE-based algorithm can expedite the evaluation of the RSC of an MG. The computational effort will be further discussed in the next section. Besides, the implementation of BESS in \textbf{Step 6} can be modified if the identified dominant random variables are located on the same branch of an MG. In such a case, the BESS located on the branch can be implemented to smooth out multiple random variables on the same branch. Also, it relaxes the assumption of having one BESS installed for each uncertainty source. Please see the results in the next section for details.  
%The algorithm to evaluate the performance of the BESS is as below: 
% $S_{b,i}$ is the set of buses inside the smoothing zone of BESS $i$ and connecting to the dominant random variables 
%\begin{enumerate}[Step 1:]
%\item Implement BESS $b_i$ at the bus where the random variable $x_i$ corresponds to. 
%\item For each time period $(m,h)$, BESS $b_i$ is scheduled to operate at $P_{B,b_i}$ determined by \eqref{equ:Pbesssmooth} to smooth the output of $P_{G,i}(x_i^{m,h})$. 
%\item Add constraints expressed in \eqref{equ:Pbesssmooth} to \eqref{equ:pADC}. 
%\item Perform Step \ref{step:generate}-\ref{step:RSC} in Section \ref{sec:calculatingRSC} to get the $RSC^{m,h}_{sm}$ after using BESS to smooth the most influential random variables. 
%\end{enumerate}

\section{Simulation Results}\label{sec:results}
\begin{figure}[htb]
    \centering
    \includegraphics[width=0.8\linewidth]{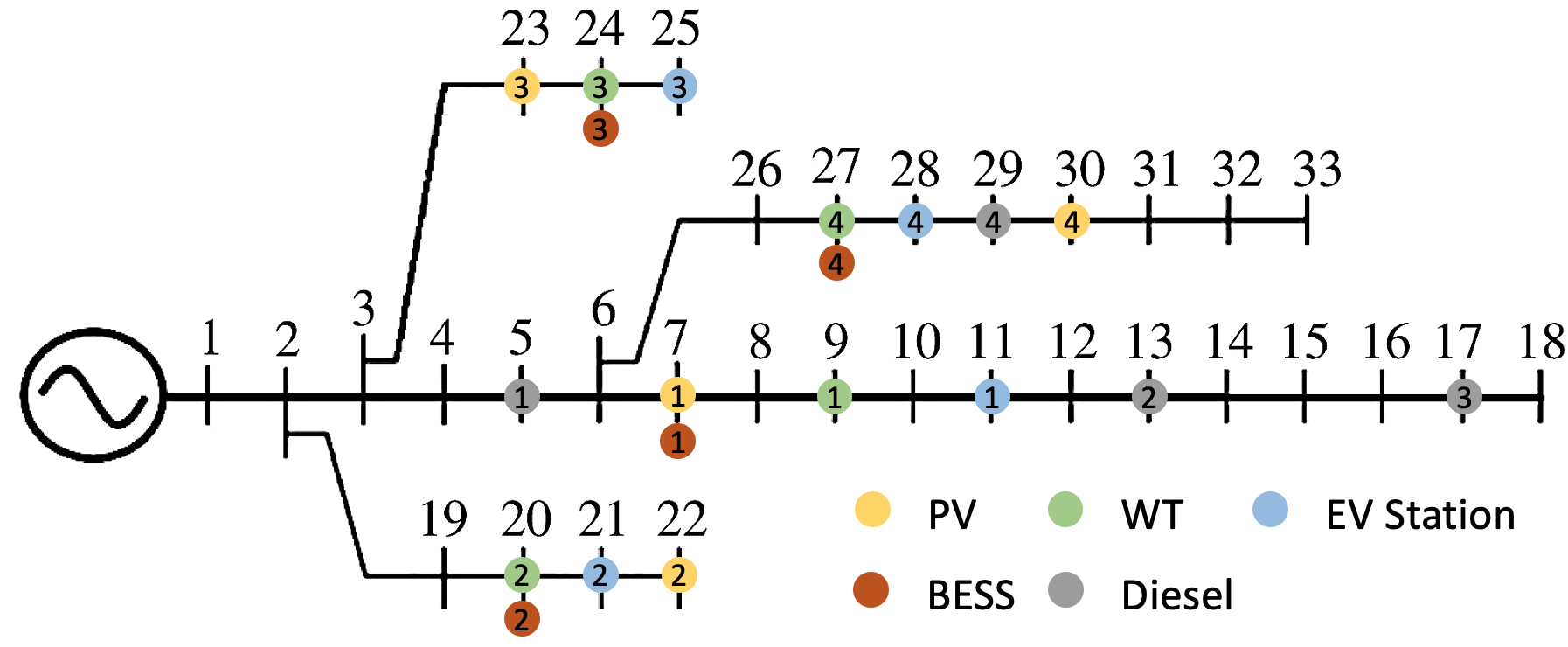}
    \caption{Diagram of the test MG.}
    \label{fig:test_grd}
\end{figure}

We applied the DDSPCE-based algorithm to assess and enhance the RSC of a modified IEEE 33-bus MG \cite{Baran1989} presented in Fig. \ref{fig:test_grd}. The MG has four 2-MW PVs with a unit power factor, four 2.25-MW WTs with a power factor of 0.85, and four 2-MW EV charging stations with a unit power factor, i.e., 12 independent random inputs in total. Besides, four 6-MW DGs with a uniform power factor of 0.93 and four 6-MW BESS with a uniform capacity of 12 MWh and unit power factor are installed in the MG. 
The data on solar radiation and wind speed is acquired from \cite{EnvironmentandClimateChangeCanada}, and the EV charging data is from \cite{Lee2019}. 
For each PV farm, the radiation set-point is 150 W/m$^2$, and the standard radiation is 2000 W/m$^2$. For each WT, the rated wind speed is 25 m/s, the cut-in speed is 4 m/s, and the cut-off speed is 40 m/s.

In the base case, we %considered no BESS in the MG 
 dispatched the four DGs simultaneously to calculate the RSC $\lambda$ of the MG. %We performed MCS with 
$N_0=250$ sample pairs were used to build the DDSPCE model. Then  %obtained the DDSPCE model for the pre-smoothing scenario, and get the distributions of RSC from 
$N_s=10,000$ samples of $\boldsymbol{x}$ were substituted to the established DDSPCE model to estimate the PDF of the RSC $\lambda$ in each time slot. %Fig. \ref{fig:distribution}. For the sake of simplicity, 
%\color{black}
Particularly, for each time slot, 
it takes 153 seconds on average to perform $N_0=250$ sample evaluations in \textbf{Step 2} and 0.178 seconds on average to obtain $N_s=10,000$ estimations of $\lambda$ in \textbf{Step 4} with Intel Core i7-8700 (3.20 GHz), 16 GB RAM. \color{black}In other words, the average time to assess RSC for one time slot by the proposed DDSPCE is about  153 seconds. \color{black}
The fast speed of the proposed method demonstrates its feasibility in online hour-by-hour RCS estimation. 
%\colorbox{yellow}
\color{black}{In contrast, 10,000 MCS take approximately 1.7 hours.}\color{black}
%\color{red} In contrast, .... (add the approximated time consumption for MC simulations with LHS, no need to run MC simulation, you may use the time of one simulation $\times$ 10,000) 

 	\begin{figure}[htb]
		\centering
		\includegraphics[width=\linewidth]{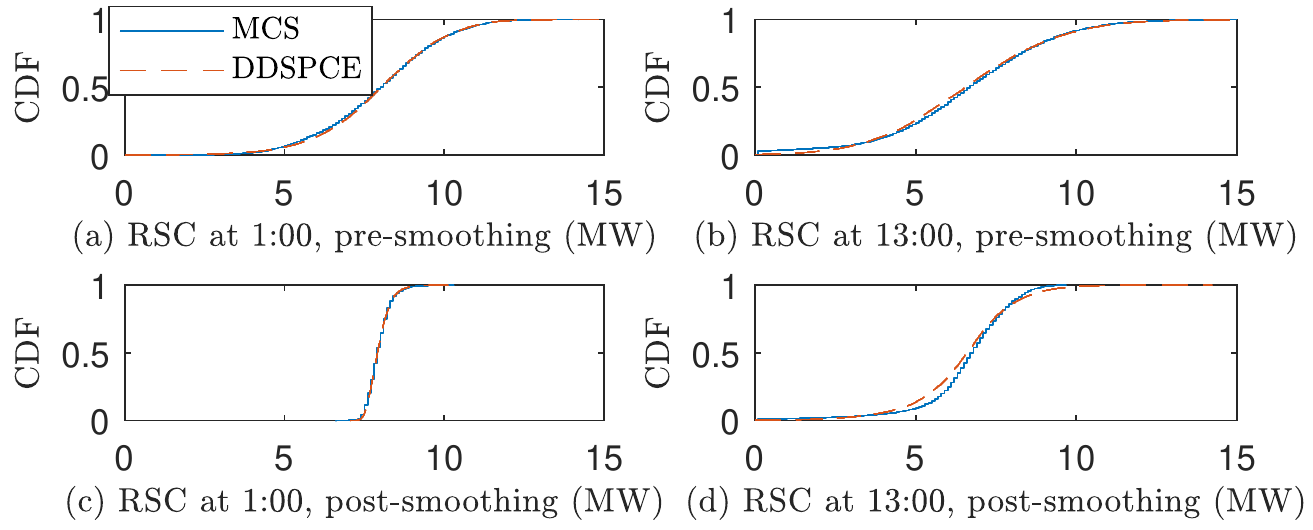}
		\caption{\color{black}Comparison between the CDF of RSC obtained from MCS and DDSPCE in scenarios (a) 1:00 pre-smoothing and (b) 13:00 pre-smoothing and (c) 1:00 post-smoothing and (d) 13:00 post-smoothing. \color{black}}
		\label{fig:revise-accuracy}
	\end{figure}

	\begin{figure*}[htb]
		\centering
		\includegraphics[width=\linewidth]{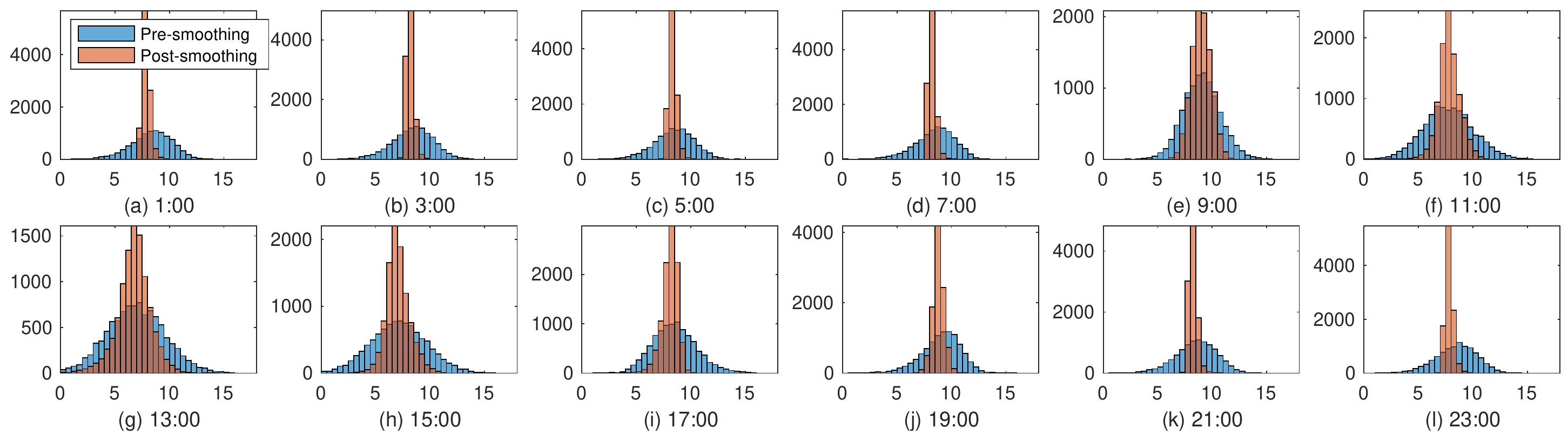}
		\caption{Distributions of pre-/post-smoothing RSC in selected time slots.}
		\label{fig:distribution}
	\end{figure*}
	
\begin{table*}[htbp]
  \centering
  \caption{Pre-/Post-smoothing RSC of the test MC in selected time slots.}
    \begin{tabular}{l|cccccccccccc}
    \toprule
    Hour & 1:00 & 3:00 & 5:00 & 7:00 & 9:00 & 11:00 & 13:00 & 15:00 & 17:00 & 19:00 & 21:00 & 23:00\\ 
    \midrule
    RSC in the base case (MW)& 4.96 & 5.05 & 5.27 & 5.13 & 6.32 & 4.02 & 2.47 & 2.85 & 5.38 & 5.71 & 5.25 & 5.14\\ 
    RSC after implementing BESS (MW)& 7.41 & 7.69 & 7.83 & 7.76 & 7.52 & 6.27 & 4.06 & 5.45 & 6.73 & 7.93 & 7.72 & 7.33\\ 
    RSC increment (MW) & 2.45 & 2.64 & 2.56 & 2.63 & 1.2 & 2.25 & 1.59 & 2.6 & 1.35 & 2.22 & 2.47 & 2.19\\
    \bottomrule
    \bottomrule
    \end{tabular}%
  \label{tab:RSC}%
\end{table*}%

	\begin{figure*}[htb]
		\centering
		\includegraphics[width=\linewidth]{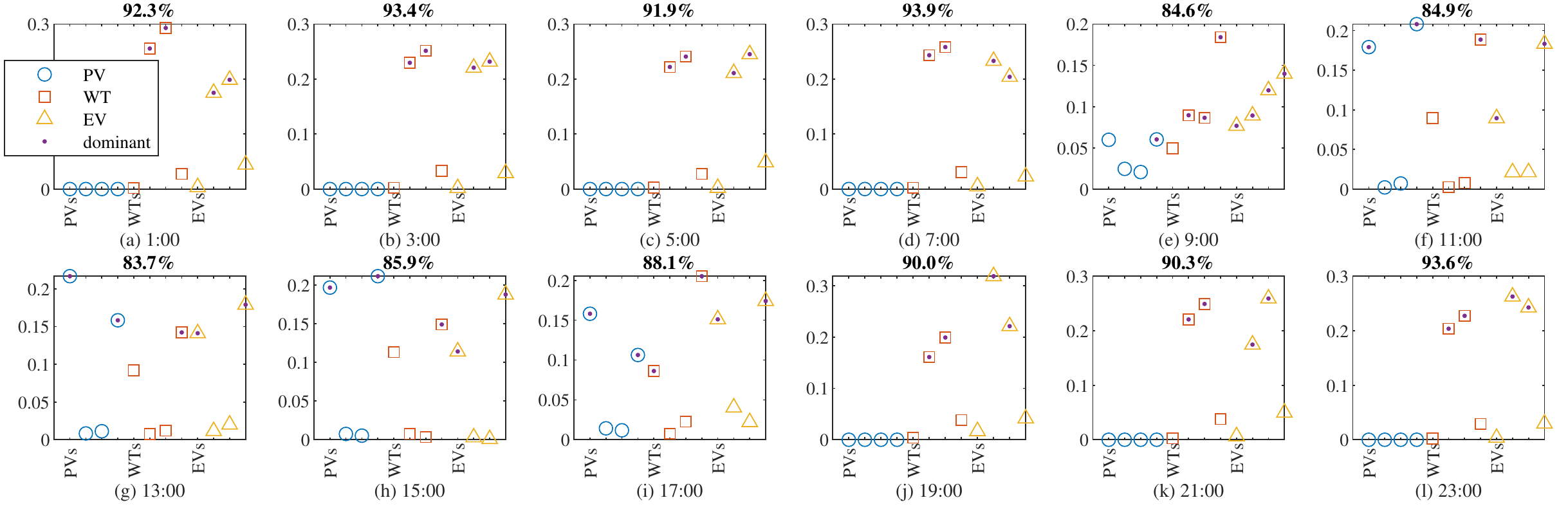}
		\caption{Sobol' index for each random input and the dominant random variables in selected time slots.}
		\label{fig:sobol_TM8}
	\end{figure*}

\color{black} First, to illustrate the accuracy of the proposed DDSPCE-based method in estimating the statistic properties of RSC, we compare the cumulative distribution functions (CDFs) of RSC obtained from MCS and those from the DDSPCE in {four} scenarios---midnight (1:00) and noon (13:00) during pre- and post-smoothing scenarios. %\colorbox{yellow}{done} \color{red}  bring the figure forward as figure 2 
\color{black} Fig. \ref{fig:revise-accuracy} shows that the estimated CDFs from the DDSPCE are always overlapping with those from the benchmark MCS, demonstrating the accuracy of the proposed DDSPCE-based method. \color{black}
Next, we present the estimated PDFs of the RSC  bi-hourly by the proposed DDSPCE \color{black} in Fig. \ref{fig:distribution}. 
The RSC with a $95\%$ confidence level RSC$_{95\%}$ in each time slot is also given in Table \ref{tab:RSC}. 
Due to the uncertainty brought by PVs in the daytime, the variance of RSC is larger than that in the night, 
which results in a smaller RSC$_{95\%}$ in the daytime as can be seen from Fig. \ref{fig:distribution} and Table \ref{tab:RSC}. Furthermore, 
the Sobol' indices and dominant influencers in each time slot are presented in Fig. \ref{fig:sobol_TM8}. The sum of Sobol' indices of dominant influencers is on the head of each subplot. 

Since some dominant influencers are adjacent on the same branch, e.g., WT 2 and EV 2 on buses 20 and 21, respectively, one BESS adjacent to them is enough to smooth them out, i.e., reduce the variance of the power output from the dominant random variables on the same branch to zero. 
The PDFs of RSC after implementing the BESS %in the post-smoothing scenario 
are presented in Fig. \ref{fig:distribution}. Compared to the PDFs of RSC in the base case, the post-smoothing ones are significantly narrower with smaller variances. As a result, the RSC$_{95\%}$ in each time slot is increased significantly, as shown in Table \ref{tab:RSC}.

\section{Conclusion}\label{sec:conclusion}
This paper proposes a DDSPCE-based method to accurately and efficiently evaluate the RSC of an MG integrating volatile RES and EVs. Moreover, the developed DDPCE model is exploited %to calculate Sobol's indices 
to pinpoint dominant uncertainty sources, based on which a scheduling method of BESS is developed to enhance the RSC of an MG.  
The proposed DDSPCE-based method, requiring no %inaccurate 
pre-assumed distributions of uncertain sources can use historical/predicted data to build the DDSPCE model efficiently online for evaluating and enhancing the hour-by-hour RSC of an MG. Simulation results in the modified IEEE 33-bus MG showed that the proposed method takes less than 3 minutes to evaluate and enhance the hourly RSC.

\bibliography{2023_GM.bib}

\begin{thebibliography}{10}
\providecommand{\url}[1]{#1}
\csname url@samestyle\endcsname
\providecommand{\newblock}{\relax}
\providecommand{\bibinfo}[2]{#2}
\providecommand{\BIBentrySTDinterwordspacing}{\spaceskip=0pt\relax}
\providecommand{\BIBentryALTinterwordstretchfactor}{4}
\providecommand{\BIBentryALTinterwordspacing}{\spaceskip=\fontdimen2\font plus
\BIBentryALTinterwordstretchfactor\fontdimen3\font minus
  \fontdimen4\font\relax}
\providecommand{\BIBforeignlanguage}[2]{{%
\expandafter\ifx\csname l@#1\endcsname\relax
\typeout{** WARNING: IEEEtran.bst: No hyphenation pattern has been}%
\typeout{** loaded for the language `#1'. Using the pattern for}%
\typeout{** the default language instead.}%
\else
\language=\csname l@#1\endcsname
\fi
#2}}
\providecommand{\BIBdecl}{\relax}
\BIBdecl

\bibitem{Majzoobi2017}
A.~Majzoobi and A.~Khodaei, ``{Application of microgrids in providing ancillary
  services to the utility grid},'' \emph{Energy}, vol. 123, pp. 555--563, mar
  2017.

\bibitem{Chiang2015}
H.~D. Chiang and H.~Sheng, ``{Available Delivery Capability of General
  Distribution Networks with Renewables: Formulations and Solutions},''
  \emph{IEEE Transactions on Power Delivery}, vol.~30, no.~2, pp. 898--905, apr
  2015.

\bibitem{Denholm2013}
\BIBentryALTinterwordspacing
P.~Denholm, M.~O'connell, G.~Brinkman, and J.~Jorgenson, ``{Overgeneration from
  Solar Energy in California: A Field Guide to the Duck Chart},'' National
  Renewable Energy Laboratory, Tech. Rep., 2013. [Online]. Available:
  \url{www.nrel.gov/publications.}
\BIBentrySTDinterwordspacing

\bibitem{Kumar2019}
A.~Kumar, N.~K. Meena, A.~R. Singh, Y.~Deng, X.~He, R.~C. Bansal, and P.~Kumar,
  ``{Strategic integration of battery energy storage systems with the provision
  of distributed ancillary services in active distribution systems},''
  \emph{Applied Energy}, vol. 253, p. 113503, nov 2019.

\bibitem{Dong2010}
Z.~Dong and P.~Zhang, \emph{{Emerging techniques in power system
  analysis}}.\hskip 1em plus 0.5em minus 0.4em\relax Springer Berlin
  Heidelberg, 2010.

\bibitem{NorthAmericanElectricReliabilityCouncil1996}
\BIBentryALTinterwordspacing
{North American Electric Reliability Council}, ``{Available Transfer Capability
  Definitions and Determination},'' North American Electric Reliability
  Council, Princeton, Tech. Rep., jun 1996. [Online]. Available:
  \url{http://www.ece.iit.edu/$\sim$flueck/ece562/atcfinal.pdf}
\BIBentrySTDinterwordspacing

\bibitem{Majzoobi2016}
A.~Majzoobi and A.~Khodaei, ``{Application of microgrids in addressing
  distribution network net-load ramping},'' \emph{2016 IEEE Power and Energy
  Society Innovative Smart Grid Technologies Conference, ISGT 2016}, dec 2016.

\bibitem{Yan2017}
X.~Yan, D.~Abbes, and B.~Francois, ``{Uncertainty analysis for day ahead power
  reserve quantification in an urban microgrid including PV generators},''
  \emph{Renewable Energy}, vol. 106, pp. 288--297, jun 2017.

\bibitem{Alharbi2015}
W.~Alharbi and K.~Raahemifar, ``{Probabilistic coordination of microgrid energy
  resources operation considering uncertainties},'' \emph{Electric Power
  Systems Research}, vol. 128, pp. 1--10, nov 2015.

\bibitem{Wang2011a}
M.~Q. Wang and H.~B. Gooi, ``{Spinning reserve estimation in microgrids},''
  \emph{IEEE Transactions on Power Systems}, vol.~26, no.~3, pp. 1164--1174,
  aug 2011.

\bibitem{Sheng2018}
H.~Sheng and X.~Wang, ``{Applying polynomial chaos expansion to assess
  probabilistic available delivery capability for distribution networks with
  renewables},'' \emph{IEEE Transactions on Power Systems}, vol.~33, no.~6, pp.
  6726--6735, nov 2018.

\bibitem{Wang2021}
X.~Wang, X.~Wang, H.~Sheng, and X.~Lin, ``{A Data-Driven Sparse Polynomial
  Chaos Expansion Method to Assess Probabilistic Total Transfer Capability for
  Power Systems with Renewables},'' \emph{IEEE Transactions on Power Systems},
  vol.~36, no.~3, pp. 2573--2583, may 2021.

\bibitem{Yu2009}
H.~Yu, C.~Y. Chung, K.~P. Wong, H.~W. Lee, and J.~H. Zhang, ``{Probabilistic
  load flow evaluation with hybrid latin hypercube sampling and cholesky
  decomposition},'' \emph{IEEE Transactions on Power Systems}, vol.~24, no.~2,
  pp. 661--667, 2009.

\bibitem{Huang2011}
J.~Huang, Y.~Xue, Z.~Y. Dong, and K.~P. Wong, ``{An adaptive importance
  sampling method for probabilistic optimal power flow},'' \emph{IEEE Power and
  Energy Society General Meeting}, 2011.

\bibitem{Liu2022}
J.~Liu, X.~X. Wang, and X.~X. Wang, ``{A Sparse Polynomial Chaos
  Expansion-Based Method for Probabilistic Transient Stability Assessment and
  Enhancement},'' \emph{IEEE General Meeting Power \& Energy Society}, pp.
  1--5, nov 2022.

\bibitem{Marelli2022}
S.~Marelli, N.~L{\"{u}}then, and B.~Sudret, ``{UQLAB USER MANUAL POLYNOMIAL
  CHAOS EXPANSIONS},'' 2022.

\bibitem{Marelli2022a}
S.~Marelli, C.~Lamas, K.~Konakli, C.~Mylonas, P.~Wiederkehr, and B.~Sudret,
  ``{UQLAB USER MANUAL SENSITIVITY ANALYSIS},'' 2022.

\bibitem{Baran1989}
M.~E. Baran and F.~F. Wu, ``{Network reconfiguration in distribution systems
  for loss reduction and load balancing},'' \emph{IEEE Transactions on Power
  Delivery}, vol.~4, no.~2, pp. 1401--1407, 1989.

\bibitem{EnvironmentandClimateChangeCanada}
\BIBentryALTinterwordspacing
{Environment and Climate Change Canada}, ``{About Ottawa (Kanata -
  Orl{\'{e}}ans)}.'' [Online]. Available:
  \url{https://ottawa.weatherstats.ca/about.html}
\BIBentrySTDinterwordspacing

\bibitem{Lee2019}
\BIBentryALTinterwordspacing
Z.~J. Lee, T.~Li, and S.~H. Low, ``{ACN-Data -- A Public EV Charging
  Dataset},'' jun 2019. [Online]. Available:
  \url{https://ev.caltech.edu/dataset}
\BIBentrySTDinterwordspacing

\end{thebibliography}
\bibliographystyle{IEEEtran}

\end{document}